\documentclass[useAMS,usenatbib,usegraphicx]{mn2e}

\title[Escape fraction of ionizing photons from galaxies]
{Escape fraction of ionizing photons from galaxies at $z=0$--$6$}
\author[A. K. Inoue, I. Iwata, J.-M. Deharveng]
{Akio K. Inoue$^{1}$\thanks{E-mail: akinoue@las.osaka-sandai.ac.jp
(AKI)}, Ikuru Iwata$^{2}$, and Jean-Michel Deharveng$^{3}$\\
$^{1}$College of General Education, Osaka Sangyo University, 3-1-1,
Nakagaito, Daito, Osaka 574-8530, Japan\\
$^{2}$Okayama Astrophysical Observatory, National Astronomical
Observatory of Japan, Kamogata, Okayama 719-0232, Japan\\
$^{3}$Laboratoire d'Astrophysique de Marseille, Traverse du Siphon, BP
8, 13376 Marseille, CEDEX 12, France}
\begin{document}

\date{Accepted Received; in original form}

\pagerange{\pageref{firstpage}--\pageref{lastpage}} \pubyear{2006}

\maketitle

\label{firstpage}

\begin{abstract}
 The escape fraction of ionizing photons from galaxies is a crucial
 quantity controlling the cosmic ionizing background radiation and the
 reionization. Various estimates of this parameter can be obtained in the
 redshift range, $z=0$--6, either from direct observations or from the
 observed ionizing background intensities. We compare them homogeneously 
 in terms of the observed flux density ratio of ionizing
 ($\sim900$ \AA\ rest-frame) to non-ionizing ultraviolet  
 ($\sim1500$ \AA\ rest-frame) corrected for the intergalactic
 absorption. The escape fraction is found to increase by an order of
 magnitude, from a value less than 0.01 at $z\la1$ to 
 about $0.1$ at $z\ga4$.
\end{abstract}

\begin{keywords}
 cosmology: observation --- diffuse radiation 
 --- galaxies: evolution --- intergalactic medium
\end{keywords}

\section{Introduction}

Although recent observations have revealed the outline of the cosmic
reionization \citep{pag06,fan06}, its detailed history and
the nature of ionizing sources are not yet fully understood. Early 
forming galaxies can be  strong ionizing sources, depending on the escape
fraction of their ionizing photons ($f_{\rm esc}$). This $f_{\rm esc}$ 
is therefore a key quantity for understanding the cosmic
reionization process but its typical (or effective) value is not yet 
clearly established
\citep[e.g.,][]{ino05}.

The goal of this Letter is to put together all existing information on
the amount of ionizing photons released by galaxies into the
intergalactic medium (IGM), both from direct observations and from
indirect estimations based on the effects of the IGM ionization itself. 
In order to make an homogeneous comparison, we introduce a quantity
derived from $f_{\rm esc}$, the {\it escape} 
flux density (Hz$^{-1}$) ratio of the Lyman continuum (LyC, 
$\lambda \sim 900$ \AA\  rest-frame) to the non-ionizing 
ultraviolet (UV, $\lambda\sim1500$ \AA\  rest-frame), 
${\cal R}_{\rm esc}$. 
This ${\cal R}_{\rm esc}$ is naturally measured by direct 
observations \citep[e.g.,][]{ste01}.
We compile all the observations of the LyC from galaxies and derive 
${\cal R}_{\rm esc}$ at $z\la3$. 
On the other hand, ${\cal R}_{\rm esc}$ can be derived  
from the ionizing background intensity inferred from IGM absorption
measurements via a cosmological radiative transfer model. 
We derive ${\cal R}_{\rm esc}$ at $z=0$--6 based on the recent reports
of the background intensities.

The cosmological parameters assumed in this Letter are 
$H_0=70.0$ km s$^{-1}$ Mpc$^{-1}$, 
$\Omega_{\rm m}=0.3$, and $\Omega_\Lambda=0.7$.

\section{Formulation}

The definition of $f_{\rm esc}$ in this Letter is 
\begin{equation}
 f_{\rm esc} = \frac{F_{\rm LyC}^{\rm esc}}{F_{\rm LyC}^{\rm int}}\,,
\end{equation}
where $F_{\rm LyC}^{\rm int}$ and $F_{\rm LyC}^{\rm esc}$ are the
intrinsic and escaping LyC flux densities, respectively. 
To obtain $F_{\rm LyC}^{\rm esc}$, we have to correct the
observed LyC flux density $F_{\rm LyC}^{\rm obs}$ for the photoelectric
absorption by the neutral hydrogen remained in the IGM: 
\begin{equation}
 F_{\rm LyC}^{\rm esc}=F_{\rm LyC}^{\rm obs}
  e^{\tau_{\rm LyC}^{\rm IGM}}\,.
\end{equation}
We may estimate $F_{\rm LyC}^{\rm int}$ from the observed UV flux
density $F_{\rm UV}^{\rm obs}$ as follows: 
\begin{equation}
 F_{\rm LyC}^{\rm int}={\cal R}_{\rm int} 
  F_{\rm UV}^{\rm obs} e^{\tau_{\rm UV}^{\rm ISM}}\,,
\end{equation}
where ${\cal R}_{\rm int}=F_{\rm LyC}^{\rm int}/F_{\rm UV}^{\rm int}$ 
is the intrinsic LyC-to-UV flux density ratio and 
$\tau_{\rm UV}^{\rm ISM}$ is the UV dust opacity in the interstellar
medium of a galaxy. Therefore, we have 
\begin{equation}
 f_{\rm esc}=\frac{{\cal R}_{\rm esc}}{{\cal R}_{\rm int}}
  e^{-\tau_{\rm UV}^{\rm ISM}}\,,
\end{equation}
and 
\begin{equation}
 {\cal R}_{\rm esc}=\frac{F_{\rm LyC}^{\rm obs}}{F_{\rm UV}^{\rm obs}}
  e^{\tau_{\rm LyC}^{\rm IGM}}\,.
\end{equation}
This ${\cal R}_{\rm esc}$ is called the LyC-to-UV escape flux density
ratio in this Letter. As found in equation (5), we can obtain 
${\cal R}_{\rm esc}$ from the observed LyC-to-UV flux density ratio 
with a correction for the IGM absorption against the LyC.
This is similar to the approach introduced by \cite{ste01}.

On the other hand, ${\cal R}_{\rm esc}$ can be obtained from the
observed ionizing background intensity through a cosmological radiative
transfer model. We describe the different aspects of this estimation in
the four following sub-sections.

\subsection{Cosmological Lyman continuum transfer}

The mean specific intensity at the observed frequency $\nu_{\rm o}$ as
seen by an observer at redshift $z_{\rm o}$ is given by 
\citep[e.g.,][]{pee93} 
\begin{equation}
 J(\nu_{\rm o},z_{\rm o})=\frac{(1+z_{\rm o})^3}{4\pi}
  \int_{z_{\rm o}}^{z_{\rm i}} dz \frac{dl}{dz}(z)
  \rho(\nu,z) e^{-\tau_{\rm eff}(\nu_{\rm o},z_{\rm o},z)}\,,
\end{equation}
where the frequency $\nu=\nu_{\rm o}(1+z)/(1+z_{\rm o})$, 
$dl/dz$ is the line element, $\rho$ is the emissivity per unit comoving
volume, and $\tau_{\rm eff}$ is the effective IGM opacity.
Although the upper limit of the integral could be infinity, we set it to
$z_{\rm i}$. The line element is 
\begin{equation}
 \frac{dl}{dz}(z)=\frac{c}{H(z)(1+z)}\,,
\end{equation}
where $c$ is the speed of light and $H$ is the Hubble parameter. 
The comoving emissivity can be written as
\begin{equation}
 \rho(\nu,z) = \rho_{\rm QSO}(\nu,z) + \rho_{\rm gal}(\nu,z)\,,
\end{equation}
where $\rho_{\rm QSO}$ and $\rho_{\rm gal}$ are 
the QSO and galactic emissivities, respectively. 
The IGM emissivity can be omitted (see \S2.3).
The effective IGM opacity is \citep[e.g.,][]{par80}
\begin{equation}
 \tau_{\rm eff}(\nu_{\rm o},z_{\rm o},z)=
  \int_{z_{\rm o}}^z dz' \int_{N_{\rm l}}^{N_{\rm u}} dN_{\rm HI}
  \frac{\partial^2 {\cal N}}{\partial N_{\rm HI} \partial z}
  (1-e^{-\tau})\,,
\end{equation}
where $\partial^2 {\cal N}/\partial N_{\rm HI} \partial z$ is 
the cloud number on a line of sight per unit redshift $z$ interval and
per unit H {\sc i} column density $N_{\rm HI}$ interval, and $N_{\rm l}$
and $N_{\rm u}$ are the lower and upper limits of the column density of
the IGM clouds. The cloud optical depth is
$\tau=\sigma_{\rm H}(\nu')N_{\rm HI}$ with $\sigma_{\rm H}$ the
hydrogen cross-section at the frequency 
$\nu'=\nu_{\rm o}(1+z')/(1+z_{\rm o})$.
The frequency dependence of $\sigma_{\rm H}$ for the LyC is assumed to
be $\propto \nu^{-3}$.

\subsection{IGM opacity}

The effective IGM opacity is determined by the cloud number function 
which can be expressed as \citep[e.g.,][]{mir90}
\begin{equation}
 \frac{\partial^2 {\cal N}}{\partial N_{\rm HI} \partial z}
  \propto {N_{\rm HI}}^{-\beta} (1+z)^\gamma\,.
\end{equation}
If we assume a single index $\beta$ $(1<\beta<2)$ for all $N_{\rm HI}$
and $z$, and $N_{\rm l} \sigma_{\rm H} \ll 1$ and 
$N_{\rm u} \sigma_{\rm H} \gg 1$, equation (9) can be approximated to 
\citep{zuo93,ino05}
\begin{equation}
 \tau_{\rm eff}(\nu_{\rm o},z_{\rm o},z) \approx
  \int_{z_{\rm o}}^{z} \Gamma(2-\beta) 
  {\cal A} [N_{\rm l} \sigma_{\rm H}(\nu')]^{\beta-1} (1+z')^\gamma dz'\,,
\end{equation}
where $\Gamma$ is the usual Gamma function, $\cal A$ is the number of
the IGM clouds with the column density between $N_{\rm l}$ and
$N_{\rm u}$ on a line of sight per unit redshift interval, and 
$\nu'=\nu_{\rm o}(1+z')/(1+z_{\rm o})$. According to \cite{wey98} and
\cite{kim02}, we assume $\beta=1.5$, $({\cal A}, \gamma)=(34.5,0.2)$ for 
$z\le 1.1$ and $({\cal A}, \gamma)=(6.3,2.5)$ for $z > 1.1$, and 
$N_{\rm l}=4.4\times10^{13}$ cm$^{-2}$. Note that we have assumed the
same number evolution along the redshift (the index $\gamma$) for
the Ly$\alpha$ forest and for the Lyman limit system (and also for the
damped Ly$\alpha$ clouds), whereas \cite{mad95} adopted different
indices for these clouds. As shown in the appendix of \cite{ino05}, the
same $\gamma$ for all clouds is compatible with the recent
observations of the Lyman limit systems by \cite{per03}.

While the IGM opacity model is used in this paper for the LyC, 
we display in Fig.~1 the opacity predictions at Ly$\alpha$ line where
comparisons with observations are possible. 
The Ly$\alpha$ cross-section is based on \cite{wie66}.
Note that the same IGM clouds contribute to the LyC and
Ly$\alpha$ opacities. Our model (solid line) shows a very good agreement
with the data at $z<5$ but a smaller opacity than the data at $z>5$.
An explanation is that the parameters of the cloud distribution
taken from \cite{kim02} who analysed the Ly$\alpha$ forest at $z<4$. 
If we assume $({\cal A}, \gamma)=(1.3,3.5)$ for $z>4$,
the model (dashed line) shows a very good agreement with the data.

\begin{figure}
 \includegraphics[width=8cm]{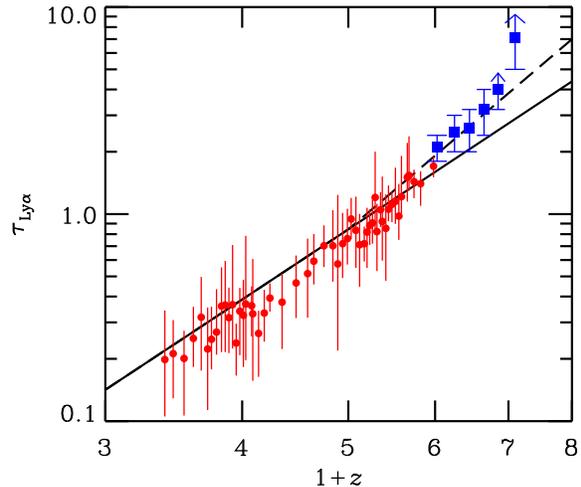}
 \caption{IGM Ly$\alpha$ opacity. The solid and dashed lines are the
 IGM Ly$\alpha$ opacities calculated by equation (11) (see the text for
 the detail). The circles and squares with error-bars are the observed
 opacity taken from Songaila (2004) and Fan et al.~(2006), respectively.}
\end{figure}

\subsection{Emissivities}

The LyC emissivity from galaxies  can be expressed as
\begin{equation}
 \rho_{\rm gal}(\nu,z)=f(\nu) {\cal R}_{\rm esc}(z)
  \rho_{\rm UV}^{\rm obs}(z)\,,  
\end{equation}
where $\rho_{\rm UV}^{\rm obs}$ is the observed galactic UV emissivity
per unit comoving volume, and $f(\nu)$ is the LyC frequency dependence
of galaxies. We assume that $f(\nu) \propto \nu^{-2}$
\citep[e.g.,][]{fio97},  independently of the redshift. 
The effect of the spectral slope on the estimated ${\cal R}_{\rm esc}$
is 15\% at most if the index is changed from 0 to $-4$.
A redshift dependence of ${\cal R}_{\rm esc}$ is introduced.

A key feature of equation (12) is that the LyC emissivity is directly
related to the observed UV emissivity with the only parameter 
${\cal R}_{\rm esc}$ that we wish to determine. In other words, 
our estimate for $R_{\rm esc}$ is free from uncertain dust correction
and from the intrinsic LyC production rate (or ${\cal R}_{\rm int}$) 
which depends on the initial mass function.

The $\rho_{\rm UV}^{\rm obs}$ values are obtained from the integration
to zero luminosity of UV luminosity functions compiled from the
literature. They are shown as a function of redshift in Fig.~2.
The error-bars indicate observational 1-$\sigma$ errors. 
For practical purpose, we express 
$\rho_{\rm UV}^{\rm obs}$ as the following functional form, 
$\rho_{\rm UV}^{\rm obs}(z)=\rho_{\rm UV}^{*}g(z)$, with
\begin{equation}
 g(z)=\cases{
  \left(\frac{1+z}{2.7}\right)^2 & ($0\le z <1.7)$ \cr
  1 & ($1.7\le z <3.0$) \cr
  \left(\frac{1+z}{4.0}\right)^{-3.5} & ($3.0 \leq z$) \cr
  }\,.
\end{equation}
The normalization adopted is $\rho_{\rm UV}^{*}=2.7\times10^{26}$ 
erg s$^{-1}$ Hz$^{-1}$ Mpc$^{-3}$ based on \cite{saw06}. 
$\rho_{\rm UV}^{\rm obs}$ of equation (13) is displayed as the solid
line in Fig.~2 and called the standard case. Because of the scatter at
$z>3$, we consider another case, the high-emissivity case, with
$g(z)=1$ for $z \ge 3$, displayed in Fig.~2 as the dotted line.

\begin{figure}
 \includegraphics[width=8cm]{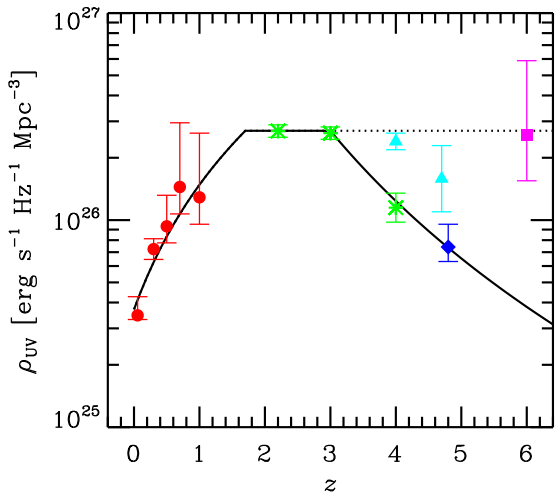}
 \caption{Observed total galactic UV emissivities per unit comoving
 volume. The circles, asterisks, diamonds, triangles, and squares  
 are taken from Schiminovich et al.~(2005), Sawicki \& Thompson (2006),
 Iwata et al.~(2006), Ouchi et al.~(2004), and Bouwens et al.~(2006),
 respectively. The solid and dotted lines are the standard and the
 high-emissivity cases assumed in the calculation, respectively.}
\end{figure}

The QSO emissivity is given by the prescription in \cite{bia01} who
obtained the LyC emissivity of QSOs based on the optical luminosity
functions of  \cite{boy00} and \cite{fan01} and an average QSO
spectrum. We here adopt a QSO spectrum parameterized by \cite{mad99}. 

The ionized IGM can contribute to a fraction of the LyC
emissivity by the recombination process \citep{haa96}. However, 
it is obviously secondary. Thus, we omit it. 
\cite{min04} suggested a significant contribution to the LyC 
by the thermal emission from hot gas shock-heated by the
cosmological structure formation. In spite of its hard spectrum,
however, the estimated hydrogen ionization rate is an order of magnitude
smaller than that reported by \cite{bol05} at $z=2$--4 and is less than
one-third of that reported by \cite{fan06} at $z=5$--6.
Thus, we also omit the thermal emission by the structure formation. 
The omission of the IGM emissivity leads to an overestimate of 
${\cal R}_{\rm esc}$ that should be less than a factor of 2.

\subsection{Estimation of the LyC-to-UV escape flux density ratio}

If we enter the IGM opacity model described in \S2.2 and the LyC
emissivities described in \S2.3 into equation (6), we can predict the
background intensity at the Lyman limit as a function of 
${\cal R}_{\rm esc}$. This latter quantity can be evaluated by
a comparison between the predicted and observed intensities. 

The observed intensities are obtained in the following ways.
At $z<4$, \cite{sco02} presented intensities at the Lyman
limit based on their observations of the QSO proximity effect. 
\cite{bol05} and \cite{fan06} estimated hydrogen ionization rates 
($\Gamma_{\rm HI}$) at $z=2$--4 and at $z=5$--6, respectively, 
based on the observed IGM Ly$\alpha$ opacity and a
cosmological hydrodynamical simulation.  The ionization rate 
$\Gamma_{\rm HI}$ is 
\begin{equation}
 \Gamma_{\rm HI} = \int_{\nu_{\rm L}}^\infty 
  \frac{4\pi \sigma_{\rm H}(\nu) J({\nu})}{h\nu} d\nu
  = \frac{4\pi \sigma_{\rm L} J_{\rm L}}{h(\alpha+3)}\,,
\end{equation}
where the subscript L means the quantity at the Lyman limit, 
$h$ is the Planck constant, and $\alpha$ is the power-law index when we 
assume a power-law background radiation ($J \propto \nu^{-\alpha}$).
Assuming $\alpha=2$, we have $J_{\rm L}/10^{-21}\,\,{\rm
erg\,\,s^{-1}\,\,cm^{-2}\,\,Hz^{-1}\,\,sr^{-1}}=\Gamma_{\rm
HI}/2.39\times10^{-12}\,\,{\rm s}^{-1}$ with 
$\sigma_{\rm L}=6.30\times10^{-18}$ cm$^2$ \citep{ost89}. 
The uncertainty on $J_{\rm L}$ resulting from the adopted $\alpha$  
is a factor of $(\alpha+3)/5$.
With this conversion  we obtain $J_{\rm L}$ from the 
$\Gamma_{\rm HI}$ values of  \cite{bol05} and \cite{fan06}. 
We should note here that \cite{fan06} determined their absolute 
$\Gamma_{\rm HI}$ values so as to fit those of \cite{mcd01} 
which are somewhat smaller than those of \cite{bol05}.

${\cal R}_{\rm esc}$ is determined by a comparison of these
observed intensities at the Lyman limit ($J_{\rm L}^{\rm obs}$) 
with the calculated theoretical intensities as
\begin{equation}
 J_{\rm L}^{\rm obs}(z)=J_{\rm L}^{\rm QSO}(z)
  + J_{\rm L}^{\rm gal}(z,{\cal R}_{\rm esc})\,, 
\end{equation}
where $J_{\rm L}^{\rm QSO}$ and $J_{\rm L}^{\rm gal}$ are the
Lyman limit intensities originating from QSOs and galaxies, respectively.
This comparison is performed by a non-parametric way from high to low
redshift. Calculating $J_{\rm L}^{\rm gal}$ (and then 
${\cal R}_{\rm esc}$) at a redshift, we use the ${\cal R}_{\rm esc}$
values at redshifts larger than the redshift.
For three different references of the observed intensities, 
we estimate ${\cal R}_{\rm esc}$ independently. The assumed initial
redshifts in equation (6) are 5.0, 5.0, and 6.0 for the data from 
\cite{sco02}, \cite{bol05}, and \cite{fan06}, respectively. This choice
does not affect the results if we take an enough high redshift because
the IGM opacity is very large at high redshift. In addition, we note
that we assumed the higher opacity case at $z>4$ (shown in Fig.~1 as
dashed line) only for the data from \cite{fan06}.

\section{Results}

The values of ${\cal R}_{\rm esc}$ obtained at different redshifts
either from direct observations or from indirect estimations (\S2.4) 
are displayed in Fig.~3.

\begin{figure}
 \includegraphics[width=8cm]{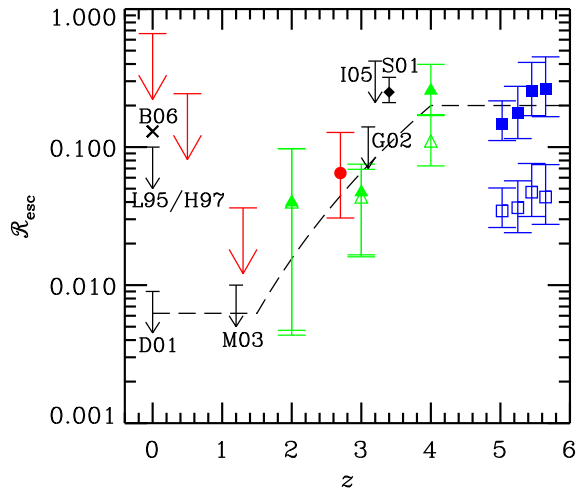}
 \caption{Escape flux density ratio of the ionizing to the non-ionizing
 UV photons. The symbols and arrows with a label are measurements and
 upper limits from direct observations: I05 (Inoue et al.~2005), S01
 (Steidel et al.~2001), G02 (Giallongo et al.~2002), M03 (Malkan et
 al.~2003), L95/H97 (Leitherer et al.~1995, Hurwitz et al.~1997), D01
 (Deharveng et al.~2001), and B06 (Bergvall et al.~2006). The circles
 and large downward arrows are the estimated value and upper limits
 based on the ionizing background intensities reported by 
 Scott et al.~(2002). Only upper limits are obtained at lower redshifts
 because the intensities from QSOs are almost equal to or exceed the
 observed intensities (see also Fig.~4). The triangles and squares are
 the estimated values based on the ionization rates reported by Bolton
 et al.~(2005) and Fan et al.~(2006), respectively; the standard
 emissivity of galaxies is assumed for the filled symbols, and the
 high-emissivity case is assumed for the open symbols. The dashed line
 is a possible fit to the evolution of ${\cal R}_{\rm esc}$ that will be
 used in Fig.~4.}
\end{figure}

Their comparison reveals an evolution of ${\cal R}_{\rm esc}$, with 
${\cal R}_{\rm esc}$ getting larger at higher redshifts. If we put
confidence in the upper limit of \cite{mal03} at $z\sim1$, which is an
average value of ${\cal R}_{\rm esc}$ for 11 star-forming galaxies and
is less affected than a single measurement by the randomness of the LyC
escape, we can obtain a possible evolution, for example, as  
\begin{equation}
 {\cal R}_{\rm esc} \propto (1+z)^5\,, 
\end{equation}
from $z=1.5$ to $z=4.0$, which is shown as the dashed line in Fig.~3. 
Since the uncertainties are still very large, this rather strong
evolution is just an example. However, we note that an evolution of 
${\cal R}_{\rm esc}$ is also found even in the
high-emissivity case (open symbols). Such an increasing  
${\cal R}_{\rm esc}$ was suggested by \cite{mei05} in his Fig.~1
although the redshift coverage is smaller than ours.

The measurement by \cite{ste01} is somewhat larger than our
estimations. This may be caused by their sample
selection bias. Indeed, their sample galaxies are taken from the bluest
quartile in the UV colour among Lyman break galaxies.
At lower redshifts, we have only upper limits, except for the
measurement by \cite{ber06}, and a large dispersion. This is probably
due to the small number of the observed galaxies. Indeed, only 7
galaxies have been observed to date \citep{lei95,deh01,ber06}. The
galaxy observed by \cite{ber06} (Haro 11) may be an exceptional one in
the local Universe as discussed in their paper.

Fig.~4 shows a comparison of the calculated ionizing intensity with the
observed ones. Assuming the evolution of ${\cal R}_{\rm esc}$ shown in
Fig.~3 as the dashed line, we obtain the dotted line for the galactic
contribution and the solid line for the total intensity. The dashed line
is the QSO contribution. The main contributor to the ionizing background
radiation changes from galaxies to QSOs at $z\sim3$.

\begin{figure}
 \includegraphics[width=8cm]{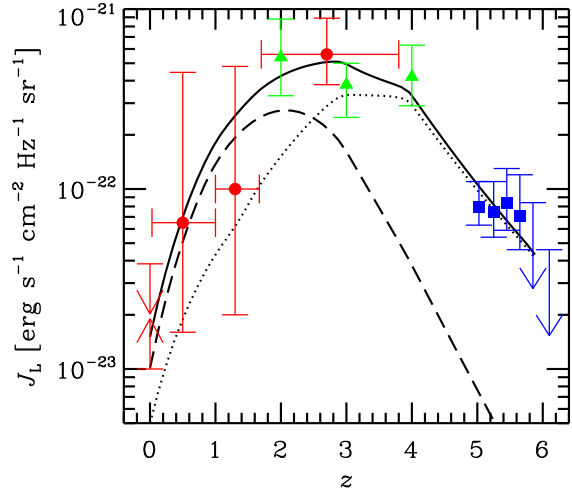}
 \caption{Lyman limit intensity of the background radiation. The circles
 and arrows at $z\simeq0$ are measurements taken from Scott et
 al.~(2002). The triangles and squares are converted from the ionization
 rates of Bolton et al.~(2005) and Fan et al.~(2006), respectively.  The
 dashed line is the intensity from the QSOs (Bianchi et al.~2001). The
 solid and dotted lines are the total and galactic intensities,
 respectively. The standard galactic emissivity and an evolution of the
 escape flux density ratio shown in Fig.~3 as the dashed line are
 assumed.}
\end{figure}

\section{Summary and discussion}

We have discussed the escape flux density ratio of the LyC to the
non-ionizing UV (${\cal R}_{\rm esc}$) in a new framework for comparing
the direct observations of the LyC from galaxies with the cosmic
ionizing background intensity. We have found a redshift evolution of 
${\cal R}_{\rm esc}$, with larger ${\cal R}_{\rm esc}$ at higher
redshifts. Here, we translate ${\cal R}_{\rm esc}$ into 
the escape fraction ($f_{\rm esc}$) of ionizing photons from galaxies.

By equation (4), we can convert ${\cal R}_{\rm esc}$ into 
$f_{\rm esc}$ if we have ${\cal R}_{\rm int}$ and 
$\tau_{\rm UV}^{\rm ISM}$. For normal stellar populations with
the standard Salpeter initial mass function and mass range 0.1--100
$M_\odot$, ${\cal R}_{\rm int}=0.2$--0.3 \citep[e.g.,][]{ino05}. 
From the observed UV slope, $\tau_{\rm UV}^{\rm ISM}=1$--2 for the Lyman
break galaxies at $z\sim3$ \citep[e.g.,][]{sha03} although the
uncertainty is very large. In this case, we have 
${\cal R}_{\rm esc}\approx f_{\rm esc}$.
If this relation is valid in the range $z=0$--6, our result means that 
$f_{\rm esc}$ increases from a value less than 0.01 at $z\la1$ to
about $0.1$ at $z\ga4$.

If we explain the evolving ${\cal R}_{\rm esc}$ by a constant 
$f_{\rm esc}$, we have two possibilities according to equation (4); 
(1) an evolving ${\cal R}_{\rm int}$ and (2) an evolving 
$\tau_{\rm UV}^{\rm ISM}$. The evolving ${\cal R}_{\rm int}$ case needs
an order of magnitude larger production rates of the LyC relative to the 
UV at $z\ga4$ than at $z\la1$. This might be realized by a
top-heavy stellar initial mass function as suggested for zero metallicity,
Population III (PopIII) stars. Very massive ($\ga100$ $M_\odot$) 
PopIII stars show almost black-body spectra with an effective
temperature about $10^5$ K \citep[e.g.,][]{bro01} and would give
${\cal R}_{\rm int}\simeq 2$. 
However, such a transition of the mass function would 
occur at a much higher redshift than $z\sim2$--4 because it is thought 
to happen at a very low metallicity like $\sim10^{-6}$
Solar value \citep[e.g.,][]{sch06}. Although \cite{jim06} proposed that 
10--30\% of stars in $z=3$--4 star-forming galaxies are massive PopIII
stars, we would need more such stars to obtain an enough large 
${\cal R}_{\rm int}$. 
The evolving $\tau_{\rm UV}^{\rm ISM}$ case needs an increment of 2--3
mag of $\tau_{\rm UV}^{\rm ISM}$ from $z\la1$ to $z\ga4$. Such a
systematic increase of $\tau_{\rm UV}^{\rm ISM}$ for galaxies at high
redshifts is not reported, whereas \cite{nol04} reported an opposite
trend for $z=2$--4 (it may be caused by a selection bias in high
redshift sample). We are therefore left with an evolution of
$f_{\rm esc}$, with a larger $f_{\rm esc}$ at a higher redshift. 

The escape of the LyC is probably a random phenomenon. Thus, the
evolving $f_{\rm esc}$ suggests an increase of the fraction of galaxies
showing a large escape. According to theoretical studies, a smaller
galaxy can show a larger $f_{\rm esc}$ because of galactic winds
\citep{fuj03} and/or champagne flows \citep{kit04}. Thus, the
evolving $f_{\rm esc}$ may suggest a decrease at higher redshift 
of the average scale of galaxies, which is consistent with the cold dark
matter scenario. The evolving $f_{\rm esc}$ also suggests a
morphological evolution of galaxies; at higher redshifts, a large
fraction of galaxies would show very disturbed structure caused
by galactic winds and/or champagne flows, yielding a large escape
of the LyC.

We note two issues which should be assessed in future. 
If there are a large number of type II QSOs at high redshifts as
suggested by \cite{mei06}, the required ${\cal R}_{\rm esc}$ may become
much smaller. If the fraction of galaxies having a low-luminosity active
galactic nucleus increases towards high redshifts as suggested by
\cite{mei05}, a scenario with an evolving ${\cal R}_{\rm int}$ without
an evolving $f_{\rm esc}$ may become possible.

Finally, we strongly encourage new direct measurements of the LyC from
galaxies to confirm/reject the evolving ${\cal R}_{\rm esc}$.
For $z\sim3$, the ground-based large telescopes are useful, and for
$z\sim1$, the {\it GALEX} satellite is appropriate.

\section*{Acknowledgments}

We thank the referee, Simone Bianchi, for comments helping us 
to improve this Letter.

\appendix

\label{lastpage}

\end{document}